\documentstyle[12pt]{article}
\newcommand{\NP}[1]{Nucl. \ Phys.}
\newcommand{\PL}[1]{Phys. \ Lett.}
\newcommand{\p}[1]{\partial}
\newcommand{\PRL}[1]{Phys.\ Rev.\ Lett. }
\newcommand{\MPL}[1] { Mod. Phys. Lett. }
\newcommand{\IJMP}[1] { Int. J. Mod. Phys. }
\begin{document}
\title{Interaction of d=2 c=1 Discrete States \\
from String Field Theory}
\author{ I.Ya.Aref'eva, \thanks{ Steklov Mathematical Institute,
Vavilov 42, GSP-1, 117966, Moscow,  E-MAIL: Arefeva@qft.mian.su} \\
P.B.Medvedev \thanks{
 Institute of Theoretical and Experimental Physics,117259, Moscow }\\ and\\
 A.P.Zubarev \thanks
{Samara Polytechnic Institute}}
\maketitle
\begin{abstract}
Starting from string field theory for 2d gravity coupled to c=1 matter
we analyze the off-shell tree amplitudes of discrete states. The amplitudes
exhibit the pole structure and we use the off-shell calculus to extract the
residues and prove that just the residues are constrained by the Ward
Identities.  The residues generate a simple effective action.  \end{abstract}

\section{Introduction}
Two-dimensional gravity coupled to a $c=1$
matter is still a subject of intense
study. The reason is that it admits non-perturbative investigations as well as
it
has a reach symmetry structure. There are two ways to describe the model.
In the first approach one deals with a matrix model and in the second
one with the continuum Liouville theory.

The characteristic  property of the model  is an appearance of discrete
states (DS) which together with tachyon states exhaust the spectrum
of physical states. This fact was
firstly recognized in the matrix model description
\cite{Gross}. The appearance of these states is rather transparent
from the Liouville point of view when the model looks like a 2d string
\cite {Pol,AZDS}. Since we deal with a string one can expect
an infinite number of higher  level states. However, since we are dealing
wiyh a 2-dimensional gauge  invariant theory we also can expect that all
these states can be eliminated by a gauge transformation. In fact due to a
nontrivial background not all the states are purely gauge. Namely, the states
with special fixed values of momentum survive in the physical spectrum
\cite{Pol,AZNSR}. Or in other words, there exist non-trivial cohomologies
of the corresponding BRST charge \cite{Lian}-\cite {IO}.

This specific nature of the spectrum provides an
enormous symmetry group \cite{KP,WZ}. The discrete  states  are spin one
currents
and they generate a $W_{\infty}$  current algebra. There are also spin
zero states generating a ground ring \cite{WGR}. It is natural to conjecture
that the existence of this algebra leads to some Ward Identities (WI) for
scattering amplitudes. This problem was addressed in \cite{Ver} where these
identities were derived for amplitudes defined by formal CFT expressions.
However, one can convince that some  of the amplitudes are ill defined  \cite
{AZDS}.  The origin of the divergences is the specific kinematics of the
"particles" living with fixed momenta only. Two-vectors of energy-momemtum
form the two-dimensional lattice  and when summing up a pair of discrete
momenta (according to energy conservation law with the background charge) one
gets the momentum of a DS again. So, all the internal lines of tree-level
amplitudes describe a propagation of  real "particles".  Hence, to obtain the
amplitude one has to deal  with  actual poles and a meaning of the WI is
rather problematic.

A natural way to expose the pole structure of a scattering amplitude is to
go off-shell. A consistent  off-shell description is provided by
string field theory (SFT).

SFT for 2d string in a nontrivial background has been proposed in \cite{AZSFT}.
As usual, the interest in SFT was motivated by the fact that
SFT is supposed to be background independent and to give a framework for
discussing
nonperturbative effects.
Now we observe that SFT appears to be a suitable tool to investigate the
symmetries of correlation functions.

In this paper we are going to use the machinery of SFT to analyse the
specific singularities of Feynman string diagrams of 2d string. The main
advantage
of SFT for the problem in question is the fact that it brings regular way to
obtain off-shell
amplitudes. A given Feynman tree diagram of SFT defines a meromorphic
function of kinematical invariants. In the usual case to obtain on-shell
tree amplitudes
one has to restrict this off-shell function on some hyperplane. In our case,
since the energies and momenta of DS are components of fixed 2d vectors,
to go on-shell
one has to sit on a fixed point and some of these points are nothing but the
actual poles.  Just these poles lead to the above mentioned divergences in
the amplitudes of CFT.

Our strategy will be the following one. We start with the general expression
for off-shell amplitudes obtained by
slightly modifying the off-shell calculation for the
critical string field theory \cite{SB}. We present
the explicit formula for the off-shell four-point
Feynman diagrams for arbitrary discrete states.
We investigate the behaviour of this formula
near the mass-shell.  For any two given external states this point is
uniquely defined by the kinematics. As one expects, this point manifests the
potential pole of the amplitude and we give the rule to calculate the
residue. In contrast to the usual case, these residues will be constants
since the value of $t$ is also fixed. In accordance with the unitarity
condition this residue must be a product of a pair of two three-point
amplitudes. The set of these three-point constants can be used to write down
an effective action such that the residues can be obtained from it. To
reflect the fact that we deal with $+$ or $-$ states numbered by pairs of
integers (half-integers) $m,n$ the effective action is a function of two sets
of variables $\Phi$ and $\bar{\Phi}$ both of which are numbered by the pairs
$(n,m)$. This effective action also seems to give the explicit expressions
for the residues of the leading poles of $n$-point off-shell tree amplitudes.
This conjecture seems to be natural as the leading singularity is fulfilled
by real "particles" in intermediate states, that brings the expression for
the residue as a product of corresponding three-point amplitudes.
Turning back to the issue of the WI we realize that the proper
objects to be constrained by the WI are  just the residues and not
the correlation functions. These identities are the origin of
the symmetries of the effective action.

The paper is organised as follows. In section 2 we summarize the main
points of the model.
In section 3 we specify the scheme of SFT off-shell calculations.
In section 4 we calculate the off-shell four-point amplitudes. In section 5
we speculate on a problem of effective action and derive the WI.
\section{The model}
In this section we shall recall the model and briefly discuss some specific
features
which are essential for our investigation.

The classical action of the model looks like an action of $d=2$ string in the
Euclidean space-time with a non-zero background charge for the Liouville mode
$\phi$
\begin{equation}
\label{1}
S=\frac{1}{8\pi} \int d^2\xi \sqrt {{\hat g}}({\hat g}^{\alpha \beta}
\partial _{\alpha}X_{\mu}\partial_{\beta}X_{\mu}-Q{\hat R}\phi +
{}~ghosts~)
\end{equation}
The value of the background charge $Q$ is fixed by the usual
requirement of vanishing of the
conformal anomaly for the total matter $ X_{\mu}=(x,\phi)$
and ghost $(b,c)$ system:
\begin{equation}
\label{2}
1+c^{\phi}=26,~c^{\phi}=1+3Q^2,~\to ~Q = 2\sqrt{2}.
\end{equation}
The non-zero $Q$ together with two-dimensional kinematics make
the spectrum of the model somewhat unusual, as was discovered in
\cite{Pol,Lian}.
The straightforward way to see it is to analyze the standard Virasoro
constraints $(L_{0}-1)\vert phys.state\rangle =0, \; L_{m}\vert
phys.state\rangle=0$ for $m>0$
in the light-cone parametrization
$$L_{m}=P^{+}(m)\alpha _{m}^{-}+P^{-}(m)\alpha_{m}^{+}+
\sum _{n\ne 0,m}\alpha _{m-n}^{+}\alpha_{n}^{-},$$
\begin{equation}
L_{0}= k^{+}k^{-}+k^{+}-k^{-}+{\hat N},
\label{3}
\end{equation}
where
\begin{equation}
P^{\pm}(m)= k^{\pm}\mp (m+1),~~k^{\pm}=\frac{1}{\sqrt{2}}(k_1\pm ik_2),~~
\alpha^{\pm}_{m}=\frac{1}{\sqrt{2}}(\alpha_{m1}\pm i\alpha_{m2}).
\label{4}
\end{equation}
The constraints $L_{m}=0$ becomes drastically simplified if one analyze
the states created only  by $\alpha ^{+}_{m}$ ($\alpha ^{-}_{m}$) modes.
Let us take a state with ${\hat N}=m_0$
\begin{equation}
\vert phys.~state\rangle=(\alpha ^{+}_{-m_0}
+\sum _{m_1+m_2=m_0}C_{m_1 m_2}
\alpha ^{+}_{-m_1}
\alpha ^{+}_{-m_2}+...) \vert k^{+},~k^{-} \rangle,
\label{5}
\end{equation}
then
$$L_{m_0}\vert phys.~state\rangle=
[P^{+}(m_0)\alpha ^{-}_{m_0}
+\sum _{n>m_0}\alpha ^{+}_{m_0-n}
\alpha ^{-}_{n}]\vert phys. ~state \rangle=$$
\begin{equation}
\label{6}
=[P^{+}(m_0)+oscillator ~ terms]\vert k^{+},~k^{-} \rangle=0
\end{equation}
and the necessary condition for a state to be physical
for some specific $m=m_0$ is $P^{+}(m_{0})=0$ which fixes the value of
$k^{+}$ to be
\begin{equation} k^{+}=m_0+1 \label{7} \end{equation} The value of
$k^{-}$ is fixed by the mass-shell condition $(L_{0}-1)\vert
phys.state\rangle =0$: $k^{-}=-2$ The example of such a physical state, the
"vector" one, is \begin{equation}
\vert phys. ~ state \rangle =\alpha ^{+}_{-1}\vert 2,~
-2 \rangle .
\label{8}
\end{equation}

In a more general setting, the discrete states appear as a non-trivial
cohomology of BRST charge $Q_{BRST}$ : $Q_{BRST}\vert \psi\rangle=0 $,
$\vert \psi\rangle \ne Q_{BRST}\vert \lambda\rangle $, and they can be
classified
according to their ghost number. The nontrivial cohomology can be found
for ghost number $0,...,3$ and it can be proved that not only the states with
fixed momenta are in the spectrum but the whole spectrum is exhausted by
these states plus the tachyon with non-discrete momentum.
The physical states in the total (matter + ghosts) Fock space can be explicitly
described in two ways: by using Shur polynomials \cite{WZ} or in terms of
$SU(2)$ raising and lowering operators \cite{WZ,KP}. For our purposes the
second
description is more convenient.

The universal formula for on-shell physical states in terms of
conformal fields $Y_r$ reads \cite{WGR}
\begin {equation} 
                                                          \label {14}
Y^{\pm}_{J,n}=cW^{\pm}_{J,n},
\end   {equation} 
where
\begin {equation} 
                                                          \label {15}
W^{\pm}_{J,J-n}=\sqrt{\frac{(2J-n)!}{n!(2J)!}}
\underbrace {[H^{-},...[H^{-}}_{n},W^{\pm}_{J,J}]...],~~
\end{equation} 
and
\begin {equation} 
                                                          \label {16}
H^{-}=\oint \frac{dz}{2\pi i }e^{-i\sqrt {2}X}(z),
{}~W^{\pm}_{J,J}=e^{i\sqrt{2}JX}e^{\sqrt{2}(1\mp J)\phi}.
\end{equation} 
$J$ is a positive integer or a half integer.
For  the sake of simplicity we shall assume the "space" dimension $x$
to be  compactified, in this case  the tachyon  state will have a discrete
momentum also.
The momenta of discrete states read \cite{Pilch}- \cite {AZNSR}:
\begin{equation}
k_{\mu}=(p,-i\varepsilon)=\sqrt{2}(n,-i(1\mp J))
\label{9}
\end{equation}
where $$J=0,\frac{1}{2},1,\frac{3}{2},2,...;~~n=-J,-J+1,...,J.$$
\section   {The Scheme of the Off-shell Calculations}

The strategy and all the necessary tools for off-shell calculations
in SFT for usual critical strings were developed in a series of papers by
Samuel et al. The interested reader can find the details in the
original manuscripts \cite{SB}. Here we adopt this calculus for the case
of the critical string in a non-trivial background.

The action for 2d SFT is taken to be of the usual Witten type and
the amplitudes are computed using perturbation theory.
With each Feynman graph  is associated a string configuration $R_{\tau}$.
External strings are semi-infinite rectangular strips of width $\pi$.
The $i$-th  internal string propagator is a strip of length $\tau _i$ and
width $\pi $. The interaction glues the strips
in pairwise manner.

The external states located at points $w_{i}$ are represented by vertex
operators $Y_r=cW_r$. For any fixed set of the off-shell states $Y_r$-s
it is possible
to obtain the the amplitude but, to treat the states of a general form
by using off-shell conformal methods
one has to guarantee that $Y_r$-s are conformal fields of definite
conformal dimension $\Delta_r$. So we have to specify the
proper rule for going off-shell.
Taking $Y_{J,n} $ in the form
\begin{equation}
Y^{\pm}_{J,n}=cV_{J,n}e^{\sqrt{2}(1\mp J)\phi}
\label{17}
\end{equation}
we can relax the mass-shell condition by simply substituting:
$\sqrt{2}(1\mp J)\phi  \to \varepsilon \phi$ for some parameter
$\varepsilon $. The fields $Y^{\varepsilon}_{J,n}=
cV_{J,n}e^{\varepsilon \phi} $
will be conformal fields with  conformal dimension
\begin{equation}
                                                          \label {18}
\Delta (\varepsilon) =-1+J^2-\frac{1}{2}\varepsilon ^{2}+\sqrt{2}\varepsilon
\end{equation}
which goes to zero on-shell ($\varepsilon = \sqrt{2}(1\mp J)$)

The contribution to a particular Feynman graph is
\begin{equation}
A_N=(\prod_{i=1}^{N-3}\int _{0}^{\infty}d\tau _{i})<Y_1(w_1)Y_2(w_2)
\oint dw'_{1}b(w'_{1})~.~.~.~Y_N(w_N)>_
{R_{\tau}},
                                   \label {10}
\end{equation}
where the correlation function is taken on string configuration.

Although the world-sheet action for the first-quantized
string is quadratic the complicated geometry of the string
configuration makes the computation non-trivial. The key idea is to
conformaly map the string configuration to the upper half-plane. The map
$\rho (z)$ which takes the half-plane into the string configuration
for the $N$-point tree diagram was given by Giddings and Martinec \cite{GM}.

The correlation function  is a product of two factors $\langle
...\rangle_{R_{\tau}}= $ $\langle ... \rangle ^{x,\phi}_{R_{\tau}}
\langle ... \rangle ^{bc}_{R_{\tau}} $
which can be calculated separately. The on-shell condition is
irrelevant for the $b-c$ factor, hence one obtains the standard result
\begin{equation}
\label{11}
<~...~>^{bc}_{R_{\tau}}\prod _{i=1}^{N-3}d\tau_{i}=
(z_1-z_2)(z_1-z_N)(z_2-z_N)\prod _{r=1}^{N}(\frac{d\rho}{dz}e^{-\rho}
)^{-1}_{z_r} \prod _{i=1}^{N-3}dz_i,
\end{equation}
where
$z_r$ are the asymptotic positions
on the real axis of $N$ external states.
The $x-\phi$ factor transforms under the mapping $\rho$ into:
\begin{equation}
\label{12}
<~...~>^{x,\phi}_{R_{\tau}}=\prod _{r=1}^{N}(\frac{d\rho}{dz}e^{-\rho})^
{-\Delta _r+1}_{z_r}\langle \prod_{r=1}^{N}W_r(z_r)\rangle .
\end{equation}
Note, that for on-shell states all the $\Delta$-s are equal to zero.
Collecting the factors (\ref{11}) and (\ref{12}) together one achieves for the
amplitude (\ref{10})
\begin{equation}
\label {13}
A_N=(\prod_{i=1}^{N-3}\int dz_{i})<Y_1(z_1)Y_2(z_2)W_3(z_3)
{}~.~.~.~W_{N-1}(z_{N-1})Y_N(z_N)>_{z-plane}(\prod _{r=1}^{N}e^{N^{rr}_{00}
\Delta _{r}}).
\end{equation}
where $N^{rr}_{00}$ are the coefficients of the Neuman functions.

\section{Four-Point Off-Shell Amplitude.}

As it has been mentioned in the Introduction,
because of the specific nature of allowed momenta one falls into a
puzzling situation as soon as the calculation of amplitudes for DS is
performed.  Namely, let us consider, for example, an s-channel four point
amplitude with the following values of particle's momenta $k_1=(\sqrt{2},0),~
k_2=(-\sqrt{2},0),~k_3=(\sqrt{2},0),~k_4=(-\sqrt{2},-i2\sqrt{2})$
which satisfy the shifted energy-momentum conservation law
in presence of the background charge: $\sum p_i =0$,$ \sum \varepsilon_i =Q$.
It appears to be natural to define the invariant variable $s$
in the presence of the background charge by the following formula
\begin{equation}
s=(k_1+k_2)^2-
(\varepsilon _1+\varepsilon _2)^2
+2\sqrt{2}
(\varepsilon _1+\varepsilon _2)
\label{210}
\end{equation}
Hence for this configuration the value of $s$ is also fixed to be
$s=0.$
We are sitting on the pole, the amplitude is obviously divergent
and this is the typical case for the theory.

In this section we derive the  expression for the residues  of the
intermediate state poles for general  $s$-channel four-point amplitude:
\begin{equation}
\label {19}
A_4=\int dz_3\langle Y_1(z_1)Y_2(z_2)W_3(z_3)
Y_4(z_4)\rangle(\prod _{r=1}^{4}e^{N^{rr}_{00}
\Delta _{r}}).
\end{equation}
Here we have omitted the $r,n$ indices for $Y$-s and labelled
them by the numbers of states.
The positions of the  points $z_i$ on the real axis are fixed by the
parameter $\alpha$ of the Giddings mapping $\rho(z)$ as follows
$z_2 =-z_3 =\alpha ~~z_1 =-z_4 =1/\alpha$

To convert the variable $\alpha$ to the Koba-Nielsen variable
$x$ one maps the upper half-plane  into itself by using $SL(2,R)$
mapping
\begin{equation}
\label {20}
x=\frac{\alpha^2 -1}{\alpha^2 +1}\frac{\alpha z+1}{\alpha z-1}
\end{equation}
The three points $z_1 ,~z_2,~ z_4$ are mapped to $\infty ,~1,~0$ as it must be
and the  final result of straightforward but tedious calculations, reads
\begin{equation}
\label {221}
A^{s}_4=\lim _{x_1 \to \infty }\int _{1/2}^{1}dx
\langle Y_1(x_1)Y_2(1)W_3(x)
Y_4(0)\rangle
\end{equation}
$$x^{2\Delta _1}x^{\frac{1}{2}(\Delta _3
+\Delta _4 -\Delta _1 -\Delta _2)}(1-x)^{\Delta _2 +\Delta _3}
[\frac{\kappa (x)}{2}]^{\sum _{r=1}^{4}\Delta _r}.$$
Here we exploited the fact that $N_{00}^{11}=N_{00}^{44}=\ln \kappa/\alpha ,~~
N_{00}^{22}=N_{00}^{33}=\ln \kappa\alpha$,
where $\kappa$ is a regular function in the region $1/2\le x \le 1$,
($\sqrt{2}-1 \ge\alpha \ge 0$), nonzero at the point $x=1$.

To analyze the pole structure of (\ref{221}) note that the poles correspond
to the divergences of the integral on the upper  limit or, in other words,
to terms in the integrand having factors of $1-x$ to negative powers.
Such factors come from the explicit factor of
$(1-x)^{\Delta _2 + \Delta _3}$ and from possible contractions of the
pair
$Y_2 (1)W_3(x)$ .
One can present the OPE for these two operators in the form
\begin{equation}
\label{22}
Y_2(1)W_3(x)\sim (1-x)^{2n_2n_3-\varepsilon _2\varepsilon _3}
(1-x)^R,
\end{equation}
where the first factor originates from the product of the exponents
and the second one - from the contraction of the Shur polynomials.
It is important to stress that $R$ is an integer. Hence, we can
present the amplitude in the form
\begin{equation}
\label{23}
A^{s}_4=\int _{1/2}^{1}dx
(1-x)^{2n_2 n_3-\varepsilon _2\varepsilon _3+\Delta _2 +\Delta
_3-R}F(x),
\end{equation}
where $F(x)=F(x;~s_r,~n_r,~\varepsilon _r) $ is a regular function at $x\sim
1 $.

Moreover, according to the adopted definition of the invariant variable
$s$ (\ref{210}) we have
\begin{equation}
                                                         \label{24}
2n_2 n_3-\varepsilon _2\varepsilon _3+\Delta _2 +\Delta
_3=\frac{1}{2}s-2,
\end{equation}
so, expanding $F(x)$ in the Taylor series
$$F(x)=\sum _{i=1}^{\infty}F_i(1-x)^i  $$
we get the final expression for $A^{(s)}_4$
\begin{equation}
\label{25}
A^{(s)}_4=\sum _{i=1}^{\infty}\frac{1}{\frac{1}{2}s-R-1+i}
(\frac{1}{2})^{\frac{1}{2}s-n-1+i}F_i.
\end{equation}

Now, we turn to extract some information from eq.(\ref{25}). Note, that
for every given on-shell states 2 and 3 $s$ is an even number:
\begin{equation}
\label{26}
s=2[(n_2+n_3)^2-(J_2+J_3)^2+2(J_2+J_3)+4]=s_0
\end{equation}
Hence, there is one and only one pole for $i_0 =R+1-\frac{1}{2}s_0$ in
$A^{(s)}_4$ and the
$F^{on-shell}_{i_0} $
is the residue (maybe equal to zero) in
this pole. Moreover, to find the residue there is no need to develop
the off-shell calculations, it is sufficient to extract from the
on-shell correlation function
\begin{equation}
\label{27}
\langle Y_1(\infty)Y_2(1)W_3(x)Y_4(0)\rangle
\end{equation}
the coefficient of the term $\sim (1-x)^{-1}$. It will be precisely
the $F^{on-shell}_{i_0}$.

As it is known the total s,t-channel amplitude is a sum of two Feyman
diagrams of SFT. The first diagram is presented below and the second one is
obtained by the cyclic permutation of the external states: $(1234)\to
(4123)$.  The residue of the corresponding t-pole obviously will be given by
the corresponding term in the expansion of eq.(\ref{27}) with permuted labels
of vertex operators.

\section   {The Effective Lagrangian}

As it was proposed by Klebanov and Polyakov \cite{KP} the model can be
described by its effective action. In this section we are going to connect the
concept of effective action with the scattering amplitudes.

The above discussion of the
four-point amplitude (see also \cite{AZDS}) shows that it is natural to
associate an effective action with the leading singularities of the
amplitudes. Indeed, using the OPE for $Y_2 (1)$ and $W_3 (x)$
\begin{equation}\label{100}
Y^+_{J_2 ,n_2}(1)W^+_{J_3 ,n_3}(x)=\frac{1}{1-x}f^{J_2 +J_3 -1,n_2 +n_3}_{J_2
n_2 ,J_3 n_3}Y^+_{J_2 +J_3 -1,n_2 +n_3}(1) + regular~terms
 \end{equation}
 the expression for the residue can be presented in
the form
\begin{equation}\label{101}
\tilde{A_4}=ResA_4 \vert_{s=s_0}=f^{J'n'}_{J_2 n_2, J_3 n_3}f^{J_1
n_1}_{J'n', J_2 n_2} \end{equation} the structure constants of OPE were
calculated in \cite{WGR,KP} $$f^{J_3n_3}_{J_1 n_1,J_2 n_2}=\delta_{J_3
,J_1+J_2-1}\delta_{n_3 ,n_1+n_2} \tilde{f}_{J_1 n_1,J_4 n_4}=$$
\begin{equation}\label{R}
\delta_{J_3 ,J_1+J_2-1}\delta_{n_3 ,n_1+n_2}
\frac{\tilde{N}(J_3m_3)}{\tilde{N}(J_1m_1)\tilde{N}(J_2m_2)}(J_2m_1-J_1m_2).
\end{equation}
Here $\tilde{N}(Jm)$  is a normalisation factor.

This result (eq.(\ref{101})) can be reproduced by a simple effective
field theory which
describes an
interaction of two independent fields $\Phi _{J,n}$ and $\bar{\Phi} _{J,n}$
with indices $J\geq 0,~ -J\leq n\leq J$ having a trivial propagator.
 The Lagrangian has the form
\begin {equation} \label {321}
{\cal L}(\Phi,\bar{\Phi }) =\sum _{a}\bar{\Phi} _{a}\Phi ^{a} +g\sum
_{a,b,c}\Phi ^{a}\Phi
^{b}\bar{\Phi }_{c}f^{b}_{ab}.
 \end   {equation}
 where $\Phi ^{a}= \Phi_{s,n}$ and $\bar{\Phi }_{a}= \bar{\Phi} _{s,-n}$ and
\begin {equation} \label {422}
f_{ab}^{c}=<Y^{+}_{a}Y^{+}_{b}Y^{-}_{c}>
\end   {equation}
It seems natural that the Lagrangian (\ref{321}) correctly reproduces the
coefficients of the leading singularities for all the n-point tree
amplitudes by a simple one to one correspondence between graphs of the SFT
and the ones of the effective theory. There is no rigorous proof at
our disposal but there are some indirect arguments in favour of this
hypothesis. In the previous paper \cite{AZDS} it was shown that if a special
regularization is adopted then the resedues of the leading poles (with respect
to the regularization parameter) are given by the effective Lagrangian
(\ref{321}). Following this interpretation of the
effective theory we conclud that there are no reasons to add to the
effective action the quartic and higher order terms if we deal with the tree
amplitudes.

Now, let us discuss the issue of Ward Identities.
The symmetry properties of on-shell amplitudes
were considered in  \cite {Ver}. The identities were derived by using the
usual contour deformation trick. Namely,
inserting the zero-dimensional charge
\begin {equation} 
                                                          \label {w1}
Q^{\pm}_{s,m}=
\oint \frac{dz}{2\pi i} W^{\pm}_{s,m}(z),
\end   {equation} 
in the correlation function, changing the contour of integration
and using the OPE's of $W^{+}_{s,m}$ and $Y^{\pm}_{s_i,m_i}$
one gets the WI.
However this consideration is rather formal, since some of amplitudes are
ill defined, as it was explained above.
The well defined objects are off-shell amplitudes and it seems natural to
search for some WI for these amplitudes.
However there is an essential obstacle to do this as it is impossible to
apply the contour deformation argument to off-shell amplitudes.
This originates from the fact
that it is  impossible to define an action of the non-local operator
(\ref {w1}) on conformal fields of non-integer or non-half-integer
 dimension. So, WI for off-shell amplitudes of SFT cannot
be derived in this way.

To solve this puzzle
let us recall that we have the expression for residues of amplitudes in terms
of on-shell correlations functions in eq.(\ref {27}). Inserting the operator
(\ref {w1}) in eq. (\ref {27}) we get the relation between corresponding
correlations functions $$\tilde{f}_{sm,s_1m_1} \langle
Y^{+}_{s_{1}+s-1,m_{1}+m}(\infty)Y^{+}_{s_{2},m_{2}}(1) W^{+}_{s_3,m_3}(x)
Y^{-}_{s_4,m_4}(0)\rangle\pm $$ $$
\tilde{f}_{sm,s_2m_2}
\langle Y^{+}_{s_{1},m_{1}}(\infty)Y^{+}_{s_{2}+s-1,m_{2}+m}(1)
W^{+}_{s_3,m_3}(x)Y^{-}_{s_4,m_4}(0)\rangle\pm $$
$$\tilde{f}_{sm,s_3m_3}\langle Y^{+}_{s_{1},m_{1}}(\infty)
Y^{+}_{s_{2},m_{2}}(1)
W^{+}_{s_3+s-1,m_3+m}(x)Y^{-}_{s_4,m_4}(0)\rangle \pm $$
\begin {equation} 
                                                          \label {w5}
\tilde{f}_{sm,s_4m_4}\langle Y^{+}_{s_{1},m_{1}}(\infty)
Y^{+}_{s_{2},m_{2}}(1)
W^{+}_{s_3,m_3}(x)Y^{-}_{s_4+s-1,m_4+m}(0)\rangle     =0
\end   {equation} 
It is  evident that this relation is also true for
the terms proportional to $(1-x)^{-1}$ in expansion of eq.(\ref{w5})
which define the corresponding residues. So, the proper objects to be
constrained
by the WI are the residues and the identities read
$$
\tilde{f}_{sm,s_1m_1}
\tilde {\cal A}((s_{1}+s+1,m_{1}+m)^{+}(s_{2},m_{2})^{+},(s_{3},m_{3})^{+}
(s_4,m_4)^{-})\pm
$$
$$\pm
\tilde{f}_{sm,s_2m_2}
\tilde {\cal A}((s_{1},m_{1})^{+}(s_{2}+s+1,m_{2}+m)^{+},(s_{3},m_{3})^{+}
(s_4,m_4)^{-})\pm  $$
$$\pm
\tilde{f}_{sm,s_3m_3}
\tilde {\cal A}((s_{1},m_{1})^{+}(s_{2},m_{2})^{+},(s_{3}+s+1,m_{3}+m)^{+}
(s_4,m_4)^{-})\pm  $$
\begin {equation} 
                                                          \label {w6}
\pm \tilde{f}_{sm,s_4m_4}
 \tilde {\cal A}((s_{1},m_{1})^{+}(s_{2},m_{2})^{+},(s_{3},m_{3})^{+}
(s_4-s+1,m_4+m)^{-})=0.
\end   {equation} 

Coming back to the effective action (\ref {321})
it is evident that relations (\ref {w6}) should represent a symmetry
of the effective Lagrangian. This symmetry does exist. Namely,
the Lagrangian (\ref {321}) is invariant under infinite
number of infinitesimal transformations
\begin {equation} 
                                                          \label {324}
\delta _{c}\Phi ^{a}= f ^{~~a}_{cb}\Phi ^{b},~~
\delta _{c} \bar{\Phi }_{a}= f _{ac}^{~~b}\bar{\Phi }_{b},
\end   {equation} 

\section   {Concluding Remarks}

In summary, we have argued that off-shell
scattering amplitudes of 2d open string  discrete states
exhibit the pole structure and the residues are described by the
effective Lagrangian with rich symmetry structure.
We have presented the rigourous proof of these statements for four-point
open string tree amplitudes. The N-point case needs more detaled analysis.

Our effective action contains only cubic terms. This does not agree the with
 Klebanov and Polyakov conjecture \cite {KP} that the effective action is
 nonpolynomial. However, it  is possible that loops will destroy its
 cubic character.

\end{document}